\documentclass{Preprint}

\usepackage[colorlinks,urlcolor=blue,linkcolor=blue,citecolor=blue]{hyperref}

\usepackage{upmath}
\usepackage{multirow}
\usepackage{booktabs}
\usepackage{mathtools}
\usepackage{wasysym}
\usepackage{amssymb}% http://ctan.org/pkg/amssymb
\usepackage{pifont}% http://ctan.org/pkg/pifont

\jvol{XX}
\jnum{XX}
\paper{8}
\jmonth{March}
\jname{Preprint}
\pubyear{2024}

\begin{document}

%Poisoning the ML supply chain

\title{Machine Learning Security against Data Poisoning: Are We There Yet?}
\author{Antonio Emanuele Cinà}
\affil{DIBRIS, University of Genova, Via Dodecaneso, Genova, Italy}
\author{Kathrin Grosse}
\affil{ENAC VITA, \'{E}cole Polytechnique F\'{e}d\'{e}rale de Lausanne, Station 18, Lausanne, Switzerland}
\author{Ambra Demontis}
\affil{DIEE, University of Cagliari, Via Marengo, Cagliari, Italy}
\author{Battista Biggio}
\affil{DIEE, University of Cagliari, Via Marengo, Cagliari, Italy}
\author{Fabio Roli}
\affil{DIBRIS, University of Genova, Via Dodecaneso, Genova, Italy}
\author{Marcello Pelillo}
\affil{DAIS, Ca' Foscari University of Venice, Via Torino, Venice, Italy}
\markboth{Machine Learning Security
against Data Poisoning}{Machine Learning Security
against Data Poisoning}
\begin{abstract}
The recent success of machine learning (ML) has been fueled by the increasing availability of computing power and large amounts of data in many different applications. However, the trustworthiness of the resulting models can be compromised when such data is maliciously manipulated to mislead the learning process.
In this article, we first review poisoning attacks that compromise the training data used to learn ML models, including
attacks that aim to reduce the overall performance, manipulate the predictions on  specific test samples, and implant backdoors in the model. We then discuss how to mitigate these attacks using basic security principles, or by deploying ML-oriented defensive mechanisms. 
We conclude our article by formulating some relevant open challenges which are hindering the development of testing methods and benchmarks suitable for assessing and improving the trustworthiness of ML models against data poisoning attacks.
\end{abstract}

\maketitle
\chapterinitial{ Introduction}
Suppose you call one of your company suppliers to understand why they stopped sending you emails about promotions. The supplier replies that they continue to send their promotion as usual and invite you to check the spam.
The supplier was right! The emails ended up in your spam folder, together with other communications from that company. 
This could have happened not by accident, but instead as a result of fraud, in which an evil competitor ensures that the email client marks any email from the victim company as spam. 
To this end, this malicious company could flood you with spam also containing the victim company's name - until your machine learning-based spam filter associates this benign name with the property ``spam'', thus trashing any future promotions. 
This scenario is an instance of a machine learning security threat called data poisoning, described already in 2008 by Nelson et al.~\cite{Nelson08SpamFilter}. Under this setting, malicious users may cause failures in ML systems (e.g., spam filters) by tampering with their training data, thereby posing real concerns about their trustworthiness. 
Several sources confirm that poisoning is already carried out in practice~\cite{Cina2022Survey}.
For example, Microsoft's chatbot Tay\footnote{\url{www.theguardian.com/technology/2016/mar/26/microsoft-deeply-sorry-for-offensive-tweets-by-ai-chatbot}} was designed to learn language by interacting with users, but instead learned offensive statements. 
Alternatively, a group of extremists submitted wrongly-labeled images of portable ovens with wheels tagging them as \emph{Jewish baby strollers} to poison Google's image search.\footnote{\url{www.timebulletin.com/jewish-baby-stroller-image-algorithm/}}

In addition to data poisoning, other attacks are threatening the reliability and robustness of ML systems~\cite{Cina2022Survey,ENISA}.%\footnote{Complementary to \cite{Cina2022Survey}, we discuss (i) the interplay between data poisoning attacks, the trustworthiness of machine learning, and AI regularization, and (ii) the Standard Security Principles that can be implemented to safeguard ML systems against poisoning attacks.}
 Evasion attacks have been conceived to force the victim's model to output wrong predictions at test time. For example, a malicious user can craft printable stickers that can force a classifier to misclassify a stop sign, posing different security concerns for user safety and self-driving industry. Privacy attacks have been devised to extract private information about the target system (via model stealing), %its users (via model extraction), 
 or its sensitive training data (via membership inference), thus undermining the system's intellectual property or users' privacy. 
The potential harm that evasion, privacy, and poisoning attacks can cause, along with other AI-related risks, have led the European Union (EU) to publish the EU AI Act~\cite{EU2019Trustworthy}, to start defining proper policies for AI trustworthiness.
These regulations require AI systems to provide not only accurate but also human-aligned decisions, which follow the principles of being explainable, fair, robust, and accountable.

Unfortunately, the road toward developing trustworthy AI/ML systems is paved with many obstacles. In particular, it is not only a problem of designing the algorithms right, since data plays a crucial role too. As Gary McGraw says, ``data matters just as much as the rest of the technology, probably more''. While data can be a strength for AI/ML models, it may also be their most vulnerable Achilles' heel. Accordingly, Kumar et al.~\cite{kumar2020adversarial}, have revealed that data poisoning is, in fact, considered the most feared threat faced today by companies that work with machine learning.
Within this work, we give a high-level overview of data poisoning attacks and recently-proposed mitigations. 
On the attack side, we study how one can compromise the ML training phase by exploiting specific vulnerabilities of AI/ML models, which is useful to provide a systematic evaluation procedure for AI/ML robustness against data poisoning. On the defense side, we investigate novel detection and sanitization strategies developed on both data and models to mitigate the impact of potential poisoning attacks. 

In the remainder of this article, we categorize the main attack scenarios that enable practical poisoning attacks against classification ML models, discuss the corresponding poisoning attacks, and how to mitigate them, following the categorization presented in~\cite{Cina2022Survey}.
While the work in~\cite{Cina2022Survey} provides an in-depth technical survey on data poisoning attacks, we focus more here on (i) the interplay between data poisoning attacks, the trustworthiness of machine learning, and AI regulation, and on (ii) also presenting more standard security principles that can be implemented to protect ML systems against poisoning attacks. 

We conclude by discussing the current limitations and open issues hindering the development of trustworthy AI/ML models against data poisoning attacks and by identifying promising future research directions. Although we do not thoroughly cover evasion and privacy attacks in detail here, we refer the reader to the European Union Agency for Cybersecurity (ENISA) report~\cite{ENISA}% \footnote{\url{https://www.enisa.europa.eu/publications/securing-machine-learning-algorithms}}
, which offers a complementary overview of the ML threat landscape. 

\begin{figure*}[t]
    \centering
    \includegraphics[width=1\textwidth]{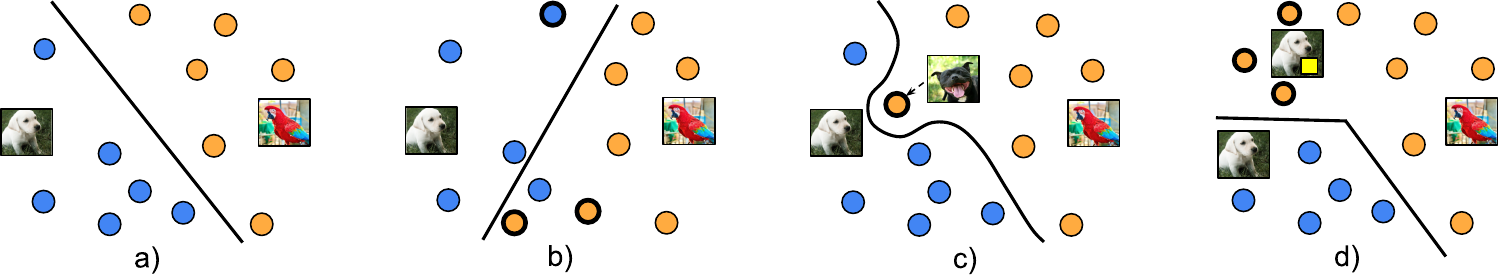}
    \caption{Conceptual figure for pristine training (a), indiscriminate (b), targeted (c), and backdoor (d)  poisoning attack. The influence of poisoning attacks leads the classifier to change its decision boundary (black line) to meet the attacker's goal.}
    \label{fig:poisoningattack}
\end{figure*}
\section{Poisoning Attacks on Machine Learning}

\begin{figure*}[t]
    \centering
    \includegraphics[width=1\textwidth]{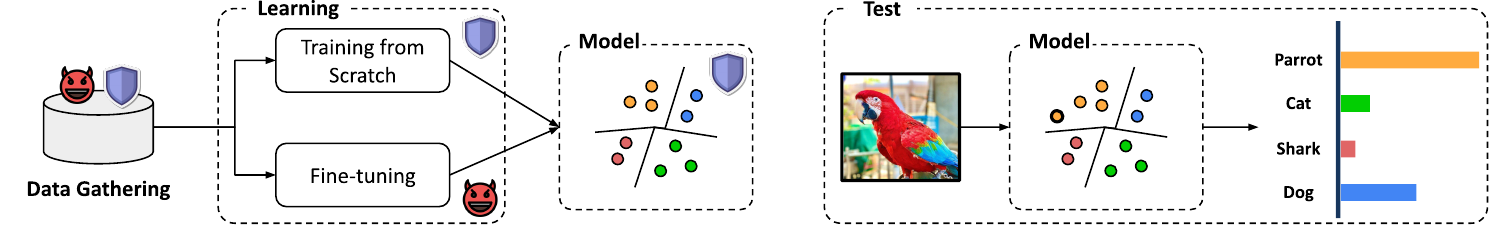}
    \caption{Training (left) and test (right) phase for a machine learning model and potential stages where a malicious user can mount a poisoning attack and where the victim user can take defensive actions.}
    \label{fig:ml_pipeline}
\end{figure*}

When developing a machine learning algorithm, the first step is to collect data, which should be ideally collected and labeled in a controlled and safe environment. However, this is a time-consuming and expensive task that not all organizations and individuals can afford. Sometimes, data is therefore collected from the Internet or other untrusted sources; e.g., when building security systems, the user can download labeled data from external vendors, such as VirusTotal\footnote{\href{https://www.virustotal.com}{https://www.virustotal.com}}, for malware data annotation. The user then splits the data into training and test datasets, which should sufficiently represent the task at hand. 
Afterward, the user can train the model from scratch on the training dataset, i.e., they train the model from a random initialization of its weights to fit the function underlying the data. However, training a machine learning model that achieves satisfactory performances might require too many computational resources if the task is complex. In this case, a pretrained model can be used and partially (or entirely) fine-tuned, e.g., trained on a smaller dataset for a short amount of time. Otherwise, the user can outsource the training procedure to third-party entities. Once the learning phase is concluded, the model can accomplish the desired task. The user finally evaluates the model on data never seen during training,  i.e., test data, to assess the model performance. If the classifier's performance on this dataset is satisfactory, the user assumes it generalizes well to other data, and they deploy the model. However, as illustrated in Fig.~\ref{fig:ml_pipeline}, the data gathering and learning phases may be entry points for malicious users (red devils) to mount a poisoning attack. Analogously, there are several phases in which defenses can be implemented (blue shields). In the following section, we thoroughly discuss the different scenarios in which attackers can threaten the trustworthiness of the user's model through data poisoning attacks and the different types of poisoning attacks that can be perpetrated. 

\subsection{Threat Modeling and Attack Scenarios}~\label{sec:threat_model}
The threat model and the attack scenarios define the set of assumptions about the setting under which attackers can mount an attack. %\textcolor{black}{The attacker may aim to cause: an availability violation to compromise the normal system functionalities or an integrity violation to cause failures only under target circumstances}. 
Three main scenarios have been considered in the literature of poisoning attacks: (i) \emph{training-from-scratch}, (ii) \emph{fine-tuning}, and (iii) \emph{model-training (or outsourcing)}.

In \emph{training-from-scratch} and \emph{fine-tuning} scenarios, the attacker influences the learning phase conducted by the victim by tampering with the training data. These attacks may occur when the victim gathers the training data from external, potentially corrupted, repositories. 
The distinction between the two scenarios hinges on how the victim employs the collected data.
In \emph{training-from-scratch} scenarios, the victim uses the collected data to train the model from scratch. In the \emph{fine-tuning} setting, instead, the victim downloads a pre-trained model and uses the collected data to fine-tune it.
In this scenario, the attacker can potentially poison the model by modifying or injecting some training samples optimized to reach a specific malicious goal. However, they might lack the details required to stage such an attack, like particular aspects related to the pristine training data and the target model. 
Nevertheless, under these limited knowledge settings, the attacker can exploit the transferability property of poisoning attacks to create effective poisoning samples. Demontis et al.~\cite{Demontis19Transfer} have shown the attacker might sample a surrogate dataset from the same distribution as the training set and use it to train a surrogate model. % by looking at the output of their queries. 
The attack can then be staged against the substitute model, and the resulting poisoning samples can finally be used to attack the target model.

Conversely, in the \emph{model-training (outsourcing)} scenario, the \textit{victim user} is assumed to have limited computational resources and outsources the training process to a third-party authority. Hence, the user could rely on external fee-based services, sharing the training dataset and a description of the desired model to train (i.e., its architecture, hyperparameters, stopping conditions, etc.). Federated learning can be considered a special case of outsourcing training where the notion of third-party authority is extended to all the participating nodes that help learn a shared model in a collaborative manner~\cite{cina2022Sponge}.
Once the model has been trained and given to the user, we assume the victim has a validation phase. The validation phase serves to assess the system's usability, i.e., the model passes the assessment if the prediction accuracy on test data, never shared with the external trainer, is satisfactory. 
With the increasing demand for computational capabilities in training deep neural networks (DNNs), this scenario is attracting the interest of users and companies with a limited budget~\cite{cina2022Sponge}. 
Rather than acquiring the necessary hardware for training such models, they prefer the pay-as-you-go formula offered by external services, often lowering costs. Yet losing complete control over the learning phase and leaving it to the third-party trainer. Despite being economically convenient, this choice leads to security risks.
An untrustworthy third-party trainer or an attacker acting as a man-in-the-middle can tamper with the training procedure and provide the victim user with a model that behaves according to their goal. 
To keep their attack stealthy, the attacker must ensure that the provided model retains high prediction accuracy, making sure to pass the validation phase without suspicion from the victim user~\cite{Gu17Badnet}.

\subsection{Attacks}
\noindent Data poisoning pollutes a ML model's training data to observe or exploit the opened vulnerability at test time.
Within the given attack scenarios, three types of poisoning attacks can be perpetrated: indiscriminate, targeted, and backdoor.
We will refer to Fig.~\ref{fig:poisoningattack} throughout this section to explain how these three attacks affect ML models. Precisely, in Fig.~\ref{fig:poisoningattack}a, we show the decision boundary of the model trained on pristine data, while the adjacent figures show the effect of the three mentioned types of poisoning attacks on it.

\paragraph{Indiscriminate Poisoning Attacks.}
Indiscriminate poisoning attacks tamper with the training data to compromise the system's availability by reducing the model's prediction accuracy on test samples. In other words, the attacker forces the model to output wrong classifications for any test input, limiting its availability and reliability to legitimate users.

In Fig.~\ref{fig:poisoningattack}b, we show an example of an indiscriminate poisoning attack adopted in Biggio et al.~\cite{Biggio12PoisonSVM}. Poisoning samples (highlighted in black) are obtained by flipping the labels of training samples, i.e., from ``dog'' (blue) to ``parrot'' (orange) and vice versa. 
As a result, the boundary of the poisoned classifier (black line) is skewed compared to its clean version (see Fig.~\ref{fig:poisoningattack}a), increasing the error for future test samples. The adversary can further compromise the system usability by increasing the fraction of flipped training samples or by carefully optimizing the injected noise in the poisoning samples features~\cite{Biggio12PoisonSVM}. 
However, while effective, the first strategy is not optimal and would require a high control of training points by the attacker, thus weighing on its applicability in real applications. Moreover, a high fraction of mislabeled training samples may induce suspicion in the victim's user, who can adopt a defense mechanism against the attack. 
On the one hand, the adversary may reduce the percentage of label flips by injecting an unbounded noise in the poisoning samples. Specifically, after flipping their labels, poisoning samples are perturbed with an adversarial noise optimized to maximize the test error of the victim's model. However, crafting these optimized samples can be computationally expensive for the attacker against cutting-edge ML models.

An alternative approach has been investigated by Feng et al.~\cite{Feng19Learning2Confuse}. To enhance stealthiness and avoid detection, the attacker optimizes the noise added to the samples to be more similar to benign samples.
Such kinds of attacks are called \textit{clean-label}, as the content of the poisoning samples seems coherent with the assigned label. However, although Feng et al.~\cite{Feng19Learning2Confuse} have improved efficiency for indiscriminate attacks and improved stealthiness compared to previous work~\cite{Biggio12PoisonSVM}, its applicability in real systems is limited to hypothesized situations where the attacker can control the entire training set.

Most of the indiscriminate poisoning attacks aim at compromising the accuracy of the target models and therefore were designed within the training-from-scratch and fine-tuning scenarios, mainly because they would not pass the validation stage provided in the model-training scenario~\cite{Cina2022Survey}.
Complementary, promising research directions, traced in %Solans et al.~\cite{Solans20Fairness} 
Jagielski et al.~\cite{Jagielski2020SubpopulationDP}, and Cinà et al.~\cite{cina2022Sponge}, aim at compromising the fairness or increasing model energy-consumption and prediction latency of the poisoned models, making them compatible with the model-training scenario as they preserve model accuracy.

\paragraph{Targeted Poisoning Attacks.}
Unlike an indiscriminate attack, which aims to lower the system's availability, a targeted poisoning attack seeks to compromise the integrity of the poisoned model.
In other words, the victim's model will hold high accuracy on clean samples but causes an induced misclassification for a target sample. This characteristic makes it compatible with the model-training scenario; however, existing attacks have been limited to training-from-scratch and fine-tuning scenarios. In Fig.~\ref{fig:poisoningattack}c we illustrate an example of a targeted attack where ``parrot'' (orange) and ``dog'' (blue) samples are almost all correctly classified. Nevertheless, the decision boundary is induced by the poisoning points (black line) to open a region where a specific ``dog'' image is classified as a ``parrot''.
Moreover, as the model performs well on benign data, it is hard to understand if the model is poisoned and thus vulnerable to targeted attacks. 
Koh et al.~\cite{Koh17Influence} were the first to show how tools designed for  interpretability of ML predictions, namely influence functions, can be used to modify relevant training samples and increase the classification error on a target sample predefined by the attacker. However, this approach has the same computational complexity as the one proposed in Biggio et al.~\cite{Biggio12PoisonSVM}, and is thus limited to shallow models.
Alternatively, Shafahi et al.~\cite{Shafahi18PoisonFrog} proposed a heuristic approach, feature collision, to mount clean-label attacks. The main idea is to leverage the complexity and non-linearity of DNNs to craft clean-label poisoning samples that collide in the feature space with a target sample the attacker would like to have misclassified. At test time, the target sample is then predicted as the poisoning sample with which it collides. For example, considering Fig.~\ref{fig:poisoningattack}c, the attacker adds noise to a ``parrot'' image to make it collide, in feature space, with the target image of a ``dog'' (the black dog shown in the figure) that they would like to have misclassified as ``parrot''. 
Although this heuristic provides promising results, it can only be used in fine-tuning scenarios. If the feature extractor is updated, the collision with the target sample will be removed, and the attack will fail.

\paragraph{Backdoor Poisoning Attacks.}
Backdoor poisoning, similar to targeted poisoning attacks, cause integrity violations, meaning that the model behaves correctly for pristine samples but misclassifies certain samples. Nevertheless, backdoor poisoning attacks are more ambitious in the attacker's goal as they lead the model to misclassify any test sample containing a specific backdoor trigger.
In this sense, the trigger is the activation mechanism that forces the model to make wrong predictions. Backdoor poisoning attacks have been investigated under all three attack scenarios, i.e., model-training, training-from-scratch, and fine-tuning. Furthermore, several strategies have been developed to stage a backdoor attack. Some of them rely on data poisoning, others tamper directly with the model's weight at test time, and others alter the training loss itself~\cite{Cina2022Survey}. Below we refer to the strategies to compromise the data to carry out the attack. 
In Fig.~\ref{fig:poisoningattack}d we depict an example of a backdoor attack where the attacker uses a yellow sticker as a trigger. The attacker embeds this trigger into a small percentage of training samples and associates the presence of the trigger with the ``parrot'' class (orange) by changing the poisoning samples' labels in \textit{parrot}. Once the model has been trained on the poisoned dataset, the adversary can exploit the vulnerability by adding yellow stickers to samples, causing integrity failures of the backdoored model.
This idea of using arbitrary patterns to implant the backdoor has been first adopted by Gu et al.~\cite{Gu17Badnet} in the model-training scenario. In their paper, the authors showed how backdoored traffic-sign classifiers could deviate their predictions from stop-sign to speed-limit signs thanks to the presence of the yellow sticker. Although performed in simple contexts, an implicit assumption of this attack is that the model has enough flexibility to learn both the original and the backdoor classification task. In addition, virtually no knowledge about the target model is required, making this attack relevant in practical applications. 

However, despite the effectiveness of Gu et al.'s attack~\cite{Gu17Badnet}, novel approaches propose the adoption of invisible triggers that can overcome human inspection and more advanced defensive techniques. Specifically, as pointed out by Doan et al.~\cite{Doan21LIRA}, the patch-based triggers proposed in~\cite{Gu17Badnet}are perceptually visible; therefore, the corresponding backdoor images can be easily detected under human inspection. Moreover, repeating the same pattern in multiple images can raise suspicion and be easily spotted by existing defensive mechanisms.  
Therefore, Doan et al.~\cite{Doan21LIRA} proposed a more stealthy backdoor attack where the adversary fits a trigger generator function that outputs a distinct invisible trigger for each poisoning sample. At test time, the backdoored model will associate the presence of such triggers with the attacker's chosen class.  
Note that, being the resulting triggers invisible and sample-specific, their detection is more challenging even for an excellent human observer.  
Nevertheless, these strengths come at a high cost. It would require solving a computationally demanding optimization problem, whereas the original~\cite{Gu17Badnet} was very efficient as the attacker exploits pre-defined triggers. 
Such simple attacks are thus practical for resource or time-constrained attackers.

\section{Defenses against poisoning}
When protecting against security threats in ML systems, defenders can combine standard security principles with ML-oriented defensive mechanisms. This integration increases the system's level of protection and reduces the risk of security breaches. Traditional security measures alone may not be sufficient for ML-based systems due to their complex nature. Integrating ML-oriented defensive mechanisms enables the system to adapt to evolving threats and maintain its integrity. In summary, combining these two approaches can result in a robust and secure ML system capable of handling novel security threats.

\subsection{Standard Security Principles}
\noindent Standard security measures can be taken to reduce the risk of an ML-based system being targeted by attacks~\cite{Saltzer75DesignPrinciples}. 
We here discuss three main strategies for safeguarding ML systems against data poisoning attacks inspired by Saltzer and Schroeder's~\cite{Saltzer75DesignPrinciples} principles. From the perspective of poisoning against ML, these directions include: (i)  \textit{access control}; (ii) \textit{system monitoring}, and (iii) \textit{audit trails}.
Access control involves designing policies to organize users and their privileges when accessing the system. To prevent data poisoning attacks, only authorized and trusted users should have the privilege to modify and validate the training data. %In the case of distributed learning, 
The defender can also randomize training data collection (e.g., collect at different times and locations) among the involved entities to further reduce the risk of receiving poisoned data from potentially-untrusted ones.
System monitoring aims to design mechanisms for continuously monitoring the system and identifying which vulnerabilities the attackers exploit. Timely detection of an attack allows the defender to promptly identify the attack's cause and take subsequent measures to reduce the impact of future attacks.
Audit trails aim to keep track of all the activities and transactions in the system; e.g., checking users and the data they manipulate allows identifying malicious users in case an attack has occurred. This enables the defender to promptly exclude those users, or degrade their privileges. 

\subsection{ML-oriented Defensive Mechanisms}
\noindent Besides standard security measures, ML-based systems demand also for the development of specific defense mechanisms to protect the AI/ML model itself. With respect to data poisoning attacks, as shown in Fig.~\ref{fig:ml_pipeline}, defenses can be deployed (i) \emph{before training}, (ii) \emph{during training}, and (iii) \emph{after training}. Different knowledge and capabilities are required to defend the model at each stage, depending on the given threat modeling scenario. %as depicted in Table~\ref{fig:matching_table}.   
We would like to finally remark that defenses are, in general, attack specific, i.e., one defense mitigates one attack~\cite{steinhardt2017certified,wang2019neural}. However, there are a few counterexamples to this~\cite{hong2020effectiveness}.

\emph{Defending before training} consists of sanitizing the data before training the model. This requires access to the training data and, in some cases, also to a set of pristine, untainted data. As in the model-training (outsourcing) scenario, if the defender does not control the data, these defenses cannot be applied. 
The underlying idea of sanitization-based defenses is that poisoning samples can be removed by outlier detection techniques, as they have to be very different from samples within the same class label (see, e.g., Figure~\ref{fig:poisoningattack}b) to induce the model to learn a significantly-different decision function~\cite{steinhardt2017certified}. Nevertheless, such defenses can be circumvented. % \rebuttal{As they} remove training samples far from the others in the same class, 
For example, attackers can fool them by applying smaller perturbations to the data. Smaller perturbations will however affect the classifier less. There is thus a trade-off between attack strength and stealthiness.

\emph{Defending during training} consists of defining robust learning algorithms that bound the influence of maliciously-altered points. 
Such approaches require access to training data and the ability to alter the model. Defenses during training are thus infeasible when the learning phase is outsourced.
One promising approach in this space relies upon limiting the impact of each training point by bounding its gradient size, based on the observation that poisoning samples exhibit larger gradients to have higher influence on the learning process 
(see, e.g., Figure~\ref{fig:poisoningattack}c where a single poisoning point altered the decision function). 
Interestingly, this technique was originally proposed in the context of differential privacy to prevent information leakage from a classifier.
In this sense, the approach in Hong et al.~\cite{hong2020effectiveness} solves two different issues simultaneously, preventing poisoning attacks and increasing privacy. 
The effect of poisoning defenses on other desiderata, such as privacy, should always be considered when applying defenses. Unfortunately, the community's current understanding of these interactions is rather limited. 

\emph{Defending after training} consists of analyzing a trained model and determining if it has been poisoned or backdoored, and potentially how to clean it, preventing wrong predictions on the incoming test data. As this group is very diverse, the defender requirements in knowledge and capabilities also vary, but these defenses are applicable in all threat modeling scenarios. For example, Wang et al.'s \emph{Neural Cleanse}~\cite{wang2019neural} reconstructs the backdoor trigger from a given model. This specific reconstruction requires a small batch of clean data and access to the model's parameters.  
To explain their approach, we consider the example in Fig.~\ref{fig:poisoningattack}d of a model that misclassifies a ``dog'' with a yellow sticker as a ``parrot''. Given the model's parameters and input samples, we can search the perturbation required to classify a ``dog'' as a ``parrot''. For a model backdoored as in \cite{Gu17Badnet}, this search likely returns a yellow sticker. Instead, for an unbackdoored model, we would find a scattered change altering, for example, the color of almost all the pixels~\cite{wang2019neural}. 
Once the trigger has been discovered, there are several ways to proceed; e.g., the defender can retrain the model with correctly labeled triggered images or remove the trigger from the incoming test samples. 		
However, an attacker aware of this defense can make it harder to reconstruct the trigger. For example, the insertion of correctly labeled images with the trigger hinders the application of \emph{Neural Cleanse}~\cite{tang2021demon}. The study of such adaptive attacks is highly relevant to evaluate defenses thoroughly. Relying on the attacker not knowing how the defense works (i.e., on \textit{security by obscurity} rather than on \textit{security by design}) should only be regarded as an additional (but not substantial) protection mechanism. So far, there are only a few works using adaptive attacks in poisoning, and thus our overall understanding of existing defenses is very limited.

\section{Open Challenges and Conclusion}
In this article, we have provided a high-level overview of data poisoning attacks on machine learning, which are considered one of the main threats to the deployment of trustworthy AI models, along with possible defense mechanisms.
We firmly believe that our overview can help stakeholders understand the risks to trustworthy ML in terms of training data security, while also being compliant with the undergoing legislation efforts both from the EU and the US. This is crucial, as poisoning attacks can also undermine other trustworthiness dimensions of AI/ML models, such as fairness and reliability of the models' predictions. A paradigmatic example is posed by federated learning, where the input of several users is combined, and even a single adversarial participant may impair the resulting model via a poisoning attack.

We would like to conclude our discussion by introducing some relevant open research challenges in the area of data poisoning attacks and defenses.
The first challenge is related to the impracticality of some threat models considered for poisoning attacks in real-world application settings. 
One example includes poisoning attacks that require control over the entire training set~\cite{Feng19Learning2Confuse}, thus limiting their viability in more realistic settings where only a few training data samples are likely to be controlled. However, although some scenarios are attractive from a mere theoretical perspective, we do believe that future work should focus more on tackling use-inspired basic research questions that pose not only methodological challenges but also better reflect realistic application constraints.

The second challenge concerns the scalability of poisoning attacks against large-scale models and modern deep networks. We have seen that heuristic attacks~\cite{Feng19Learning2Confuse,Shafahi18PoisonFrog,Gu17Badnet} exhibit promising results; however, their applicability and robustness remain limited in the presence of suitable defense mechanisms. On the other hand, effective and stealthy poisoning attacks~\cite{Biggio12PoisonSVM,Koh17Influence,Doan21LIRA} require solving an expensive optimization problem that may not scale in practice. 
It is, therefore, essential to explore more effective approximations to reduce their computational complexity, as done by Huang et al.~\cite{Huang2020MetaPoison}. %in the pioneering work by Mu{\~{n}}oz{-}Gonz{\'{a}}lez et al.~\cite{Munoz17Backgrad}. 
Some promising research avenues in this direction include building on the results from other research fields, like meta-learning and hyperparameter optimization, in which more efficient techniques to solve bilevel problems involving learning algorithms are constantly developed.

The third challenge is associated with gaining a better understanding of ML defenses. As previously discussed, privacy-enhancing mechanisms can also have an alleviating effect on poisoning attacks~\cite{hong2020effectiveness}. However, our overall understanding of how poisoning interacts with privacy or other ML security threats is very limited, and requires more in-depth investigation. 
Lastly, and more pressing, is the challenge of how to soundly evaluate defenses. Ideally, defenses should be tested against adaptive attacks, i.e., attacks that are aware of the defense mechanism~\cite{tang2021demon}. Currently, there are few adaptive attacks, and thus our knowledge about the limits of existing defenses is rather narrow.
This lack of understanding severely limits how, or if, these defenses can be applied in practice, consequently undermining the trust of any applied system trained on data potentially accessible by a malicious entity.

Bringing together these research directions can inspire the design of more effective, scalable, and practical poisoning attacks and defenses.
Such a development is crucial, as it would be otherwise difficult to properly evaluate the robustness of ML models, and analyze the factors influencing these vulnerabilities, at scale, as well as in the context of different application domains.

Finally, to answer the question raised in this article's title, we can say that \textit{no, we are not there yet} when it comes to ML security against data poisoning. While research in data poisoning has highlighted potential vulnerabilities of ML models, we are still at a very early stage.
From the attack side, we need to design more efficient attacks and  benchmark methods and protocols to systematically test the robustness and reliability of ML models under realistic poisoning threats, and at an industrial level. Similarly, for defenses, we need to develop sound evaluation guidelines and investigate the relationship with other fields to obtain more reliable and effective defenses, which can be easily deployed at scale. We do believe that tackling these challenges, especially given that data poisoning is considered one of the most relevant threats by many companies~\cite{kumar2020adversarial}, remains extremely relevant to enable the development of trustworthy ML.

\section{ACKNOWLEDGMENT}
\noindent This work has been partially supported by the PRIN 2017 project RexLearn (grant no. 2017TWNMH2), funded by the Italian Ministry of Education, University and Research; by Spoke 10 ``Logistics and Freight'' within the Italian PNRR National Centre for Sustainable Mobility (MOST), CUP I53C22000720001; and by BMK, BMDW, and the Province of Upper Austria in the frame of the COMET Programme managed by FFG in the COMET Module S3AI.

\bibliographystyle{IEEEtran}
\bibliography{short-literature.bib}

\begin{IEEEbiography}{Antonio Emanuele Cinà} is an Assistant Professor in the Department of Computer Science, Bioengineering, Robotics and Systems Engineering at the University of Genova, Italy. His research interests include machine learning security, trustworthy AI, and computer vision. Contact him at antonio.cina@unige.it.
\end{IEEEbiography}

\begin{IEEEbiography}{Kathrin Grosse} is a Postdoctoral Researcher at the School of Architecture, Civil and Environmental Engineering, EPFL, Switzerland. Her research interests include adversarial machine learning, machine learning, and machine learning security in practice. Contact her at kathrin.grosse@epfl.ch.
\end{IEEEbiography}

\begin{IEEEbiography}{Ambra Demontis} is an Assistant Professor in the Department of Electrical and Electronic Engineering at the University of Cagliari, Italy. Her research interest includes adversarial machine learning, machine learning and cybersecurity. She is a Member of IEEE. Contact her at ambra.demontis@unica.it.
\end{IEEEbiography}

\begin{IEEEbiography}{Battista Biggio} is an Associate Professor in the Department of Electrical and Electronic Engineering at the University of Cagliari, Italy. His research interest includes adversarial machine learning, cybersecurity and machine learning. He is a Fellow of IEEE. Contact him at battista.biggio@unica.it.
\end{IEEEbiography}

\begin{IEEEbiography}{Fabio Roli} is a Full Professor in the Department of Computer Science, Bioengineering, Robotics and Systems Engineering at the University of Genova, Italy. His research interest includes machine learning, computer vision and cybersecurity. He is a Fellow of IEEE and IAPR. Contact him at fabio.roli@unige.it.
\end{IEEEbiography}

\begin{IEEEbiography}{Marcello Pelillo}
is a Full Professor in the Department of Environmental Sciences, Informatics and Statistics at Ca' Foscari University of Venice, Italy. His research interests include computer vision, game theory, and pattern recognition. He is a Fellow of IEEE, IAPR, and IEEE SMC. Contact him at marcello.pelillo@unive.it.
\end{IEEEbiography}

\end{document}